\documentstyle[latexsym,amsfonts,prl,aps,twocolumn,epsf]{revtex}

\newcommand{\nn}{\nonumber\\}

\newcommand{\f}[1]{\mbox{\boldmath$#1$}}
\newcommand{\fk}[1]{\mbox{\boldmath$\scriptstyle#1$}}
\newcommand{\bea}{\begin{eqnarray}}
\newcommand{\ea}{\end{eqnarray}}
 
\begin{document} 
\title{Dielectric black hole analogues}
\author{Ralf Sch\"utzhold, G\"unter Plunien, and Gerhard Soff}
\address{Institut f\"ur Theoretische Physik, Technische  Universit\"at
Dresden, D-01062  Dresden, Germany\\
Electronic address: {\tt schuetz@theory.phy.tu-dresden.de}}
\date{\today}
\maketitle
\begin{abstract} 
Alternative to the sonic black hole analogues we discuss a different
scenario for modeling the Schwarzschild geometry in a laboratory --
the dielectric black hole.  
The dielectric analogue of the horizon occurs if the velocity of the
medium with a finite permittivity exceeds the speed of light in
that medium.  
The relevance for experimental tests of the Hawking effect and
possible implications are addressed.  
\end{abstract} 

\bigskip

At present we know four fundamental interactions in physics: the
strong and the weak interaction, electro\-magnetism and gravitation.
The first three forces are described by quantum field theories whereas 
the fourth one is governed by the laws of general relativity 
-- a purely classical theory.   
In view of the success -- and the excellent agreement with
experimental data -- of the electro-weak standard model (unifying the
electro\-magnetic and the weak force at high energies) it is conjectured 
that all four interactions can be described by an underlying unified
theory above the Planck scale.
This fundamental description is expected to incorporate the four
forces as low-energy effective theories.       
Despite of various investigations during the last decades a satisfactory
and explicit candidate for this underlying law is still missing.
At present we can just consider consistently quantum fields,
e.g.~electro\-magnetism, in the presence of classical, i.e.~externally
prescribed, gravitational fields.
This semi-classical treatment is expected to provide some insight
into the structure of the underlying theory.
One of the most striking consequences of this formalism is the Hawking 
effect \cite{hawking} predicting the evaporation of black holes. 
However, this prediction is faced with a conceptual difficulty: 
Its derivation is based on the assumption that the
semi-classical treatment is valid at {\em arbitrary} scales.
But in contrast to this presumption the decomposition into a classical 
gravitational sector and a quantum field sector is valid at {\em low}
energies only.
The investigation of high-energy effects requires some knowledge about
the underlying theory including quantum gravity. 

In order to elucidate this point Unruh \cite{unruh} suggested a
scenario which displays a close similarity to that of the
Hawking effect while the underlying physical laws are completely
understood -- the sonic black hole analogue. 
These analogues are based on the remarkable observation that the 
propagation of sound waves in flowing fluids is under appropriate
conditions equivalent to that of a scalar field in 
curved space-times:
The dynamics of the fluid is governed by the non-linear Euler equation  
$\dot{\f{v}}+(\f{v}\nabla)\f{v}+{\nabla p}/{\varrho}=\f{f}$, where
$\f{v}$ denotes the local velocity of the liquid, 
$\varrho$ its density, $p$ the pressure, and $\f{f}$ the external
force density, together with the equation of continuity 
$\dot\varrho+\nabla(\varrho\,\f{v})=0$. 
Linearizing these equations around a given flow profile $\f{v}_0$ via  
$\f{v}=\f{v}_0+\nabla\phi$ the scalar field $\phi$ of the small
deviations (sound waves) satisfies the Klein-Fock-Gordon equation with  
an appropriate (acoustic) metric 
$g_{\rm ac}^{\mu\nu}(\f{v}_0,\varrho_0)$, see e.g.~\cite{unruh,visser}
\bea
\label{KFG}
\Box_{\rm ac}\phi=\frac{1}{\sqrt{-g_{\rm ac}}}
\partial_\mu\left(
\sqrt{-g_{\rm ac}}\,g_{\rm ac}^{\mu\nu}\,\partial_\nu\phi
\right)=0
\,,
\ea
with $g_{\rm ac}={\rm det}(g^{\rm ac}_{\mu\nu}$).
The acoustic horizon occurs if the velocity of the fluid exceeds the
speed of sound within the liquid. 
Many examinations have been devoted to this topic after the original 
proposal by Unruh \cite{unruh}, see 
e.g.~Refs.~\cite{visser,volovik,liberati,garay} as well as
references therein. 
E.g., in \cite{garay} the possibility of realizing an acoustic horizon 
within a Bose-Einstein condensate is addressed.
More generally, Ref.~\cite{volovik} discusses the simulation of
phenomena of curved space-times within super-fluids.

However, the sonic analogues suffer from certain conceptual
difficulties and problems:
In order to cast the wave equation of sound into the above form 
(\ref{KFG}) it is necessary to neglect the viscosity of the liquid. 
This might be justified for super-fluids, but in general not
for normal liquids.  
Since the Hawking effect is a pure quantum radiation phenomenon, 
it can be observed for the acoustic analogues if and only if
the {\em quantum} description of sound waves is adequate.
The presence of friction and the resulting decoherence destroys the
quantum effects and so violates this presumption in general. 
In addition, the assumption of a stationary, regular, and laminar 
(irrotational) flow profile at the transition from subsonic to
supersonic flow (which was used in the derivation) is questionable,
see e.g.~\cite{liberati,blackholes}.  
Finally, we would like to emphasize that the sonic analogues
incorporate the scalar (spin-zero) field $\phi$ and are not obviously
generalizable to the electro\-magnetic (spin-one) field $A_\mu$.  

In the following we shall discuss an alternative scenario -- where all
these objections do not necessarily apply.
Let us consider the following quantum system:
Within the linear approximation the excitations of a medium are
described by a bath of harmonic oscillators whose dynamics is governed
by the Lagrangian density ${\cal L}_{\Phi}$. For reasons of simplicity
we assume these degrees of freedom to be localized and not propagating.  
The linear excitations of the medium in its rest-frame are coupled
to the microscopic electric field via ${\cal L}_{\fk{E}\Phi}$. 
We might also consider a coupling to the magnetic field whereby the 
medium would possess a non-trivial permeability in addition to its
permittivity. 
The dynamics of the microscopic electro\-magnetic field itself is
governed by ${\cal L}_{\fk{E}\fk{B}}$.
Hence the complete Lagrangian density $\cal L$ of the quantum system
under consideration is determined by
\bea
\label{fundL}
{\cal L}
&=&
{\cal L}_{\fk{E}\fk{B}}+{\cal L}_{\fk{E}\Phi}+{\cal L}_{\Phi}
=
\frac12\left(\f{E}^2-\f{B}^2\right)
\nn &&
+
\sum\limits_\alpha
\f{E}\cdot\f{\chi}_\alpha\,\Phi_\alpha
+
\frac12\sum\limits_\alpha
\left(\dot\Phi^2_\alpha-\Omega^2_\alpha\Phi^2_\alpha\right)
\,.
\ea
$\f{E}(t,\f{r})$ and $\f{B}(t,\f{r})$ denote the microscopic electric
and magnetic fields, respectively. 
The excitations of the medium are described by $\Phi_\alpha(t,\f{r})$ 
where $\alpha$ labels the different vibration modes. The fundamental
frequencies of the medium are indicated by $\Omega_\alpha$.
$\f{\chi}_\alpha$ denote the vector-valued and possibly space-time
dependent coupling parameters (e.g.~dipole moments).

Now we turn to a macroscopic description via averaging over the
microscopic degrees of freedom associated to the medium,
i.e.~the fields $\f{\Phi}=\{\Phi_\alpha\}$. 
Assuming that the microscopic quantum state of the medium is properly
represented by the path-integral with the usual regular
measure ${\mathfrak D}\f{\Phi}$ we may integrate out the microscopic
degrees of freedom $\Phi_\alpha$ within this formalism.
As a result we obtain an effective action ${\cal A}_{\rm eff}$
(see e.g.~\cite{eff}) accounting for the remaining macroscopic degrees
of freedom  
\bea
\exp\left(i{\cal A}_{\rm eff}\right)=
\frac{1}{Z_{\fk{\Phi}}}
\int{\mathfrak D}\f{\Phi}
\,\exp\left(i{\cal A}\right) 
\,.
\ea
${\cal A}$ denotes the original action described in 
Eq.~(\ref{fundL}) and $Z_{\fk{\Phi}}$ is a constant normalization
factor.  
After standard manipulations (linear substitution of $\Phi_\alpha$ and
quadratic completion, see e.g.~\cite{eff}) we may accomplish the
path-integration (averaging over all field configurations) with
respect to ${\mathfrak D}\f{\Phi}$ explicitly, and finally we arrive
at  
\bea
{\cal A}_{\rm eff}
&=&
{\cal A}_{\fk{E}\fk{B}}
\nn
&& +
\frac12\sum\limits_\alpha\int d^4x\;
\f{E}\cdot\f{\chi}_\alpha
\left(\frac{\partial^2}{\partial t^2}+\Omega^2_\alpha\right)^{-1}
\f{\chi}_\alpha\cdot\f{E}
\,.
\ea
As expected from other examples of effective field theories
\cite{eff}, the 
effective action ${\cal A}_{\rm eff}$ is non-local owing to the
occurrence of the inverse differential operator $(1+\partial^2)^{-1}$.  
However, in analogy to the operator product expansion \cite{eff} 
we may expand this non-local quantity into a sum of local operators  
$(1+\partial^2)^{-1}=\sum_n (-\partial^2)^n$. 
For the propagation of photons with energies much smaller than the
fundamental frequencies $\Omega_\alpha$ of the medium only 
the lowest term ($n=0$) of this asymptotic expansion yields
significant contributions. 
Neglecting the higher oder terms we obtain the 
local and non-dispersive low-energy effective theory of the
macroscopic electro\-magnetic field in analogy 
to a heavy-mass effective theory \cite{eff}
\bea
{\cal L}_{\rm eff}={\cal L}_{\fk{E}\fk{B}}+
\frac12\f{E}\cdot(\f{\varepsilon}-\f{1})\cdot\f{E}
\,.
\ea
The second term contains the permittivity tensor $\f{\varepsilon}$ of
the medium which is introduced via
$\f{\varepsilon}=\f{1}+
\sum_\alpha\f{\chi}_\alpha\otimes\f{\chi}_\alpha/\Omega^2_\alpha$.
In contrast to the microscopic fields in Eq.~(\ref{fundL}) here 
$\f{E}(t,\f{r})$ and $\f{B}(t,\f{r})$ are understood as the
macroscopic electric and magnetic fields, respectively. 
For reasons of simplicity we assume the coupling parameters 
$\f{\chi}_\alpha$ of the medium to be homogeneously and isotropically
distributed resulting in a constant and scalar permittivity
$\varepsilon$. 
Consequently the Lagrangian density assumes the simple form 
\bea
\label{Leff}
{\cal L}_{\rm eff}=\frac12\left(\varepsilon\,\f{E}^2-\f{B}^2\right) 
\,.
\ea
As expected, this is exactly the Lagrangian of the macroscopic
electro\-magnetic field within a dielectric medium at rest 
possessing a permittivity $\varepsilon$. 
Its dynamics is governed by the macroscopic source-free Maxwell
equations   
$\f{\nabla}\cdot\f{B}=0$, $\f{\nabla}\cdot\f{D}=0$, 
$\f{\nabla}\times\f{H}=\dot{\f{D}}$, and
$\f{\nabla}\times\f{E}=-\dot{\f{B}}$ with
$\f{D}=\varepsilon\,\f{E}$ and $\f{H}=\f{B}$ (no permeability).

Obviously the Lagrangian in Eq.~(\ref{Leff}) is not manifestly
covariant.  
However, by virtue of Lorentz transformations it can be cast into
a Poincar\'e invariant form via
\bea
\label{lorg}
{\cal L}_{\rm eff}=-\frac14\,F_{\mu\nu}\,F^{\mu\nu}
-\frac{\varepsilon-1}{2}\,F_{\mu\nu}\,u^\nu\,F^{\mu\lambda}\,u_\lambda
\,.
\ea
$F_{\mu\nu}$ symbolizes the electro\-magnetic field strength tensor
containing the macroscopic electric and magnetic fields $\f{E}$ and
$\f{B}$.  
The signature of the Minkowski metric $g_{\mu\nu}^{\rm M}$ is chosen 
according to $g^{\mu\nu}_{\rm M}={\rm diag}(+1,-1,-1,-1)$.  
$u^\mu$ denotes the four-velocity of the medium and is related to its 
three-velocity $\f{\beta}$ via 
$u^\mu=(1,\f{\beta})/\sqrt{1-\f{\beta}^2}$.

If we now assume that the dielectric properties of the medium 
(strictly speaking, the coupling parameters $\f{\chi}_\alpha$)
do not change owing to a non-inertial motion 
(generating terms such as $\partial_\mu\,u_\nu$) 
we may generalize the above expression to arbitrarily space-time
dependent four-velocities of the medium $u^\mu$, 
see e.g.~\cite{quant-rad}. 

As we shall demonstrate now, the propagation of the electro\-magnetic
field in a flowing dielectric medium resembles many features of a
curved space-time, see also \cite{landau,leo,reznik}.
Introducing the effective Gordon metric 
\bea
\label{goben}
g^{\mu\nu}_{\rm eff}=g^{\mu\nu}_{\rm M}
+(\varepsilon-1)\,u^\mu\,u^\nu\,
\ea
we may rewrite the Lagrangian in Eq.~(\ref{lorg}) into a form
associated with a curved space-time 
\bea
\label{leff}
{\cal L}_{\rm eff}=-\frac14\,F_{\mu\nu}\,g^{\mu\rho}_{\rm eff}\,
g^{\nu\sigma}_{\rm eff}F_{\rho\sigma}
\,.
\ea
Inserting the expression (\ref{goben}) into the above equation
(\ref{leff}) and exploiting the normalization of the four-velocity
$u_\mu\,u^\mu=1$ together with the anti-symmetry of the tensor
$F_{\mu\nu}=-F_{\nu\mu}$ we exactly recover the original formula
(\ref{lorg}).    

Note, that the representation of the field strength tensor via the 
{\em lower} indices is related to the four-vector potential $A_\mu$ by  
$F_{\mu\nu}=\partial_\mu\,A_\nu-\partial_\nu\,A_\mu$. 
The equation of motion assumes the compact form 
$\partial_\mu(\,g^{\mu\rho}_{\rm eff}\,
g^{\nu\sigma}_{\rm eff}\,F_{\rho\sigma})=0$. 
For a rigorous identification we have to consider the associated 
Jacobi determinant $\sqrt{-g_{\rm eff}}$ accounting for the
four-volume element. Fortunately the determinant of the metric in
Eq.~(\ref{goben}) is simply given by    
$|\det(g^{\mu\nu}_{\rm eff})|=\varepsilon>1$. This positive and
constant factor can be eliminated by a scale transformation of the
distances $dx^\mu$ or -- alternatively -- by re-scaling the
four-vector potential $A_\mu$.  

In order to discuss the general properties of the Gordon metric it is 
convenient to calculate its inverse
\bea
\label{gunten}
g_{\mu\nu}^{\rm eff}=g_{\mu\nu}^{\rm M}
-\frac{\varepsilon-1}{\varepsilon}\,u_\mu\,u_\nu
\,.
\ea
The condition of an ergo-sphere $g_{00}^{\rm eff}=0$ 
(see e.g.~\cite{ellis}) is satisfied for
$\f{\beta}^2=1/\varepsilon$, i.e.~if the velocity of medium equals 
the speed of light in that medium.
For a radially inward/outward flowing medium,
i.e.~$\f{\beta}=f(r)\f{r}$ an analysis of the geodesics 
(e.g.~light rays) reveals that the ergo-sphere represents an apparent 
horizon, cf.~\cite{ellis}.  
An inward flowing medium corresponds to a black hole 
(nothing can escape) while an outward flowing medium represents a
white hole (nothing can invade), see e.g.~\cite{visser,onhawking}.
Assuming an eternally stationary flow the apparent horizon coincides
with the event horizon, cf.~\cite{ellis}.

For a stationary flow the effective metric in Eq.~(\ref{gunten})
describes a stationary space-time. However, this metric can be cast
into a static form 
$ds^2_{\rm eff}=g_{00}d\tilde{t\,}^2-d\tilde{r\,}^2/g_{00}-r^2d\Omega^2$  
by virtue of an appropriate coordinate transformation 
$dt\rightarrow d\tilde t=dt+g_{01}dr/g_{00}$ as well as 
$dr\rightarrow d\tilde r=\sqrt{g_{01}^2-g_{00}g_{11}}\,dr$. 
Whereas the former expression in Eq.~(\ref{gunten}) corresponds to the  
Painlev{\'e}-Gullstrand-Lema{\^\i}tre metric 
(see e.g.~\cite{visser,onhawking}) the latter form is equivalent to
the well-known Schwarzschild representation. 
Since the Schwarzschild metric is singular at the horizon ($g_{00}=0$)
also the transformation $t\rightarrow\tilde t$ maintains this
property. Thus the coordinate $\tilde t$ should be used with special
care.  
  
Having derived the Schwarzschild representation of the dielectric
black/white hole its surface gravity \cite{ellis} can be calculated
simply via $2\kappa=(\partial g_{00}/\partial\tilde r)_{(g_{00}=0)}$ 
and yields
\bea
\kappa=\frac{1}{1-{\beta}^2}\,
\left(\frac{\partial\beta}{\partial r}\right)_{\rm Horizon}
\,.
\ea
Independently of $\varepsilon$ it coincides -- up to the relativistic
factor $(1-\f{\beta}^2)^{-1}$ -- with the surface gravity of the 
non-relativistic sonic analogues.  
The associated Hawking temperature \cite{hawking} is (for $\beta\ll1$)  
of the order of magnitude  
\bea
T_{\rm Hawking}
=
\frac{\kappa\,\hbar\,c}{4\pi\,k_{\rm B}}
=
{\cal O}\left(\frac{\hbar\,c}{k_{\rm B}\,n\,R}\right)
\,,
\ea
where $\hbar$ denotes Planck's constant, $c$ the speed of light in 
the vacuum, and $k_{\rm B}$ Boltzmann's constant. 
$n=\sqrt{\varepsilon}$ is the index of refraction of the medium and $R$
symbolizes the Schwarzschild radius of the dielectric black/white hole. 
Hence the Hawking temperature is proportional to the speed of light
in the medium $c/n$ over the Schwarzschild radius $R$, i.e.~the
inverse characteristic time scale of a light ray propagating within
the dielectric black/white hole. 

The dielectric analogues of the Schwarzschild geometry provide a
conceptual clear scenario for studying the effects of quantum fields
(${\cal L}_{\rm eff}$) under the influence of external conditions 
($g_{\mu\nu}^{\rm eff}$) together with their relation to the
underlying theory ${\cal L}$. 
The effective low-energy description ${\cal L}_{\rm eff}$ exhibits a
horizon (one-way membrane \cite{ellis}) and the related thermodynamical
implications (Hawking radiation \cite{hawking}).
Moreover, it allows for investigating the effects of a finite
cut-off (trans-Planckian problem) induced by the microscopic theory 
${\cal L}$. 
By means of the analogy outlined below these considerations might shed
some light on the structure of quantum gravity and its semi-classical
limit since for the dielectric black holes the underlying physics is
understood.
\begin{figure}[ht]
\centerline{\mbox{\epsfxsize=7cm\epsffile{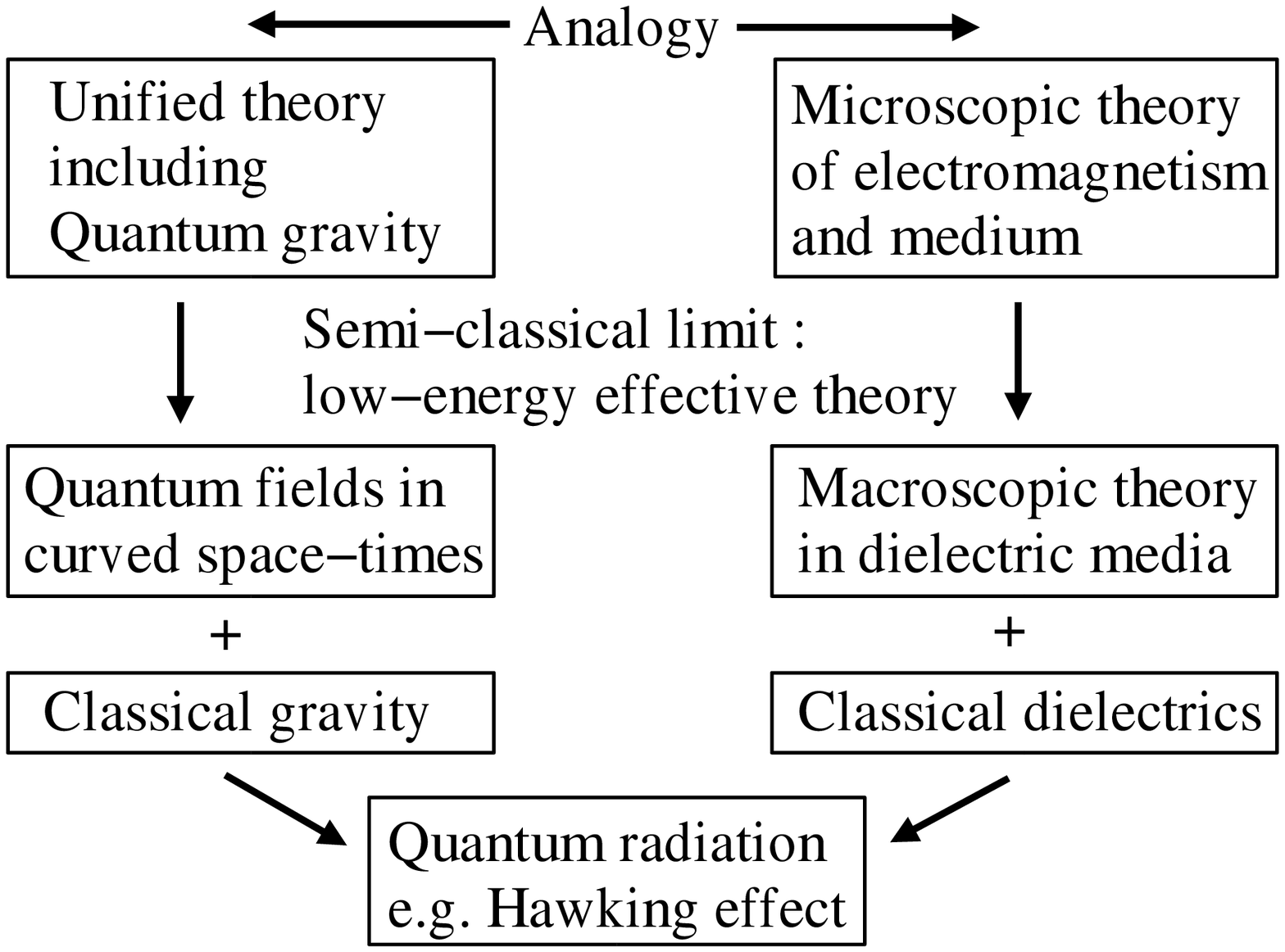}}}
\label{analogy}
\end{figure}
Apart from the considerations above there have been already a few
discussions in the literature concerning the simulation of black
holes by means of dielectric properties: 
The solid-state analogues proposed in Ref.~\cite{reznik} identify
the permittivity $\varepsilon^{ij}$ and the permeability $\mu^{ij}$ 
tensor directly with the (singular) Schwarzschild metric via  
$\varepsilon^{ij}=\mu^{ij}=\sqrt{-g}\,g^{ij}/g_{00}$.
As a result the solid-state analogues require divergences of the  
material characteristics of the medium at rest in order to simulate a
horizon ($g_{00}=0$) -- in contrast to the model described in the
present article.  
Even though such a singular behavior might be realized for the case
of a phase transition the validity of the effective theory in the
presence of these divergences is questionable.
In addition, the genuine difference between the black hole and the
white hole horizon (see e.g.~\cite{onhawking}) cannot be reproduced by
the scenario proposed in \cite{reznik}.

Apart from this model a flowing dielectric medium obeying a finite
permittivity was suggested in Ref.~\cite{leo} in order to represent a  
so-called optical black hole.
However, the Aharonov-Bohm scenario under consideration in
Ref.~\cite{leo} -- where a pure swirling of the medium is assumed --
does {\em not} exhibit a horizon and therefore {\em cannot} be
regarded as a model of a black/white hole. This objection has already
been raised in \cite{comment}, see also \cite{reply}.   
Consequently it is not possible to apply the concepts of surface
gravity and Hawking temperature.   

In contrast to the simulation of a curved space-time by a dielectric
medium it is also possible to consider the inverse identification:
E.g., in Ref.~\cite{landau} the propagation in a gravitational
background was mapped to electro\-magnetism within a medium in flat
space-time.  
However, the identification in \cite{landau} requires $g_{00}>0$ and
thus does not incorporate space-times with a horizon.
\begin{figure}[ht]
\centerline{\mbox{\epsfxsize=6cm\epsffile{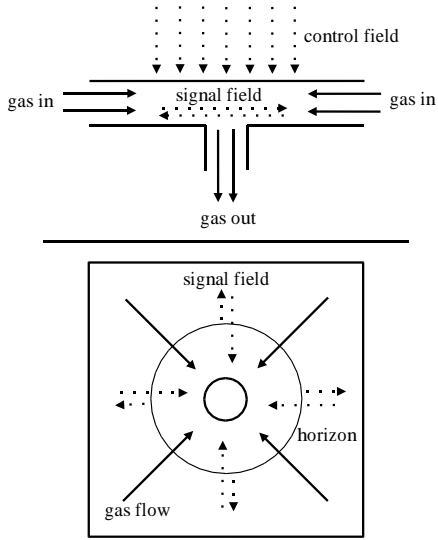}}}
\caption{Sketch of a possible experimental set-up for a dielectric
black hole in profile and top view.}   
\label{figure}
\end{figure}
The experimental realization of the dielectric analogues might indeed
become feasible in view of the recent experimental
progresses in generating ultra-slow light rays, 
see e.g.~\cite{kash,budker,fleisch,koch,phillips}.
These experiments exploit the phenomenon of the electro\-magnetically
induced transparency in order to slow down the velocity of light
in atomic vapor.
Although the microscopic theory of this phenomenon is not properly
described by the Lagrangian in Eq.~(\ref{fundL}) it generates similar 
macroscopic effects.
If an appropriate control field (a laser) acts on the medium the group
velocity of the signal field perpendicular to the control
field reaches an order of magnitude of some meters per second --
i.e.~even below the speed of sound.
In view of the non-destructive nature \cite{phillips} of the
propagation, i.e.~the absence of loss, dissipation, and the resulting
decoherence, one might imagine the observability of quantum field
theoretical effects -- e.g.~the Hawking effect -- in an advanced
experiment. 

Very roughly, an experimental set-up for a dielectric black hole as
depicted in the figure might be conceivable.

Apart from the experimental challenge of simulating a black hole
within a laboratory the scenario discussed in this article may help to 
understand the structure of the underlying theory including quantum
gravity.
%
%
\addcontentsline{toc}{section}{References}

\end{document}